\documentclass[a4paper,11pt]{article}
\usepackage{pos}

\title{Chromo-electric screening length in 2+1 flavor QCD}
\ShortTitle{Chromo-electric screening length}

\author*[a]{Peter Petreczky}
\author[b]{Sebastian Steinbei{\ss}er}
\author[c]{Johannes Heinrich Weber}

\affiliation[a]{Physics Department, Brookhaven National Laboratory\\
  Upton, NY, 11973, USA}

\affiliation[b]{Physik Department, Technische Universit\"at M\"unchen, \\
James-Franck-Str. 1, D-85748 Garching b. M\"unchen, Germany}

\affiliation[c]{Institut f\"ur Physik $\&$ IRIS Adlershof, Humboldt-Universit\"at zu Berlin, Zum Gro{\ss}en Windkanal 6, D-12489
Berlin, Germany}

\emailAdd{petreczk@bnl.gov}
\emailAdd{sebastian.steinbeisser@tum.de}
\emailAdd{dr.rer.nat.weber@gmail.com}

\abstract{We study the Polyakov loop as well as the correlators of real and imaginary parts
of the Polyakov loop in 2+1 flavor QCD at finite temperature. We use hypercubic (HYP) 
smearing to improve the signal in the lattice calculations and to obtain reliable results
for the correlators at large distances. From the large distance behavior of the correlators
we estimate the chromo-electric screening length to be $(0.38-44)/T$. Furthermore, we show that
the short distance distortions due to HYP smearing do not affect the physics of interest.}

\FullConference{%
 The 38th International Symposium on Lattice Field Theory, LATTICE2021
  26th-30th July, 2021
  Zoom/Gather@Massachusetts Institute of Technology
}

\begin{document}
\maketitle

\section{Introduction}
Polyakov loop and Polyakov loop correlators are central for understanding deconfinement
and color screening in gauge theories \cite{Kuti:1980gh,McLerran:1981pb} (see
also Refs. \cite{Petreczky:2004xs,Bazavov:2020teh} for a historic reviews). They are related to
the free energy of a static quark, $F_Q(T)$ and the free energy of a static 
quark anti-quark pair, $F_{Q\bar Q}(r,T)$ 
separated by some distance $r$. In SU(N) gauge theories the Polyakov loop and
Polyakov loop correlators are order parameters for deconfinement. 
In particular, the free energy of a static quark is infinite in the confined phase.
In gauge theories
with fundamental fields, e.g. QCD this is no longer the case, $F_Q(T)$ is 
finite even at low temperatures because the static quark is screened by
the dynamical quarks in the medium. At sufficiently low temperatures $F_Q(T)$ is
related to the binding energy of static-light hadrons \cite{Bazavov:2013yv}.
The free energy of a static quark anti-quark pair decays exponentially at large 
distances. Above the crossover temperature, the screening mass, that governs
the exponential decay of the correlators is proportional to the temperature, up to
possible sub-leading logarithmic corrections.
This means that the chromo-electric screening length that is the inverse of the screening mass
becomes shorter and shorter with increasing temperature.

Extracting the screening mass or equivalently the chromo-electric screening length from
the Polyakov loop correlator on the lattice is challenging  because of the poor signal to
noise ratio. The problem is particularly severe in QCD with light dynamical quarks, and most
of the lattice calculations of the screening masses have been performed in pure gauge theory
\cite{Kaczmarek:1999mm,Digal:2003jc,Datta:2002je}. It has been suggested to improve the signal
to noise ratio for the Polyakov loop correlator by using smeared gauge links when calculating
the Polyakov loop  \cite{Borsanyi:2015yka}.
In this contribution we report  on lattice QCD calculations of Polyalov loop
and Polyakov loop correlators using HYP smearing \cite{Hasenfratz:2001hp}.

We performed calculations of the Polyakov loop and Polyakov loop correlators in 2+1 flavor
QCD using highly improved staggered quark (HISQ) action \cite{Follana:2006rc}
with physical strange quark mass, $m_s$ and light quark masses $m_l=m_s/20$, which correspond
to the pion mass of $161$ MeV in the continuum limit. The gauge configurations used in this study
have been generated by the HotQCD and TUMQCD collaborations 
\cite{HotQCD:2014kol,Bazavov:2016uvm,Bazavov:2017dsy,Bazavov:2018wmo,Bazavov:2019qoo}
on lattices with temporal extent $N_{\tau}=4,~6,~8,~10,~12$ and $16$ and the ratio of spatial
to temporal extent (aspect ratio) $N_{\sigma}/N_{\tau}=4$. 
For $N_{\tau}=4$ and $6$ we also consider the aspect ratio $N_{\sigma}/N_{\tau}=6$.
Our calculations span a temperature range from 110 MeV
to 6 GeV. We use 1, 2, 3 and 5 steps of HYP smearing. On the finest lattices we also use 8 steps
of HYP smearing.

\section{Polyakov loop with HYP smearing}
\begin{figure}
\includegraphics[width=7cm]{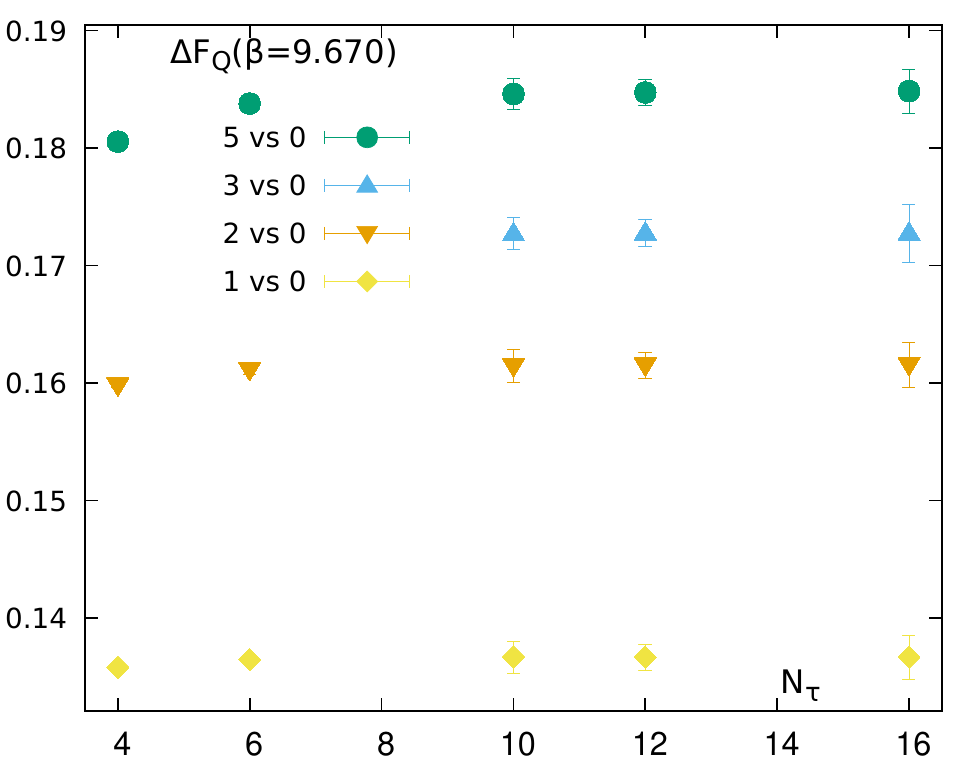}
\includegraphics[width=7cm]{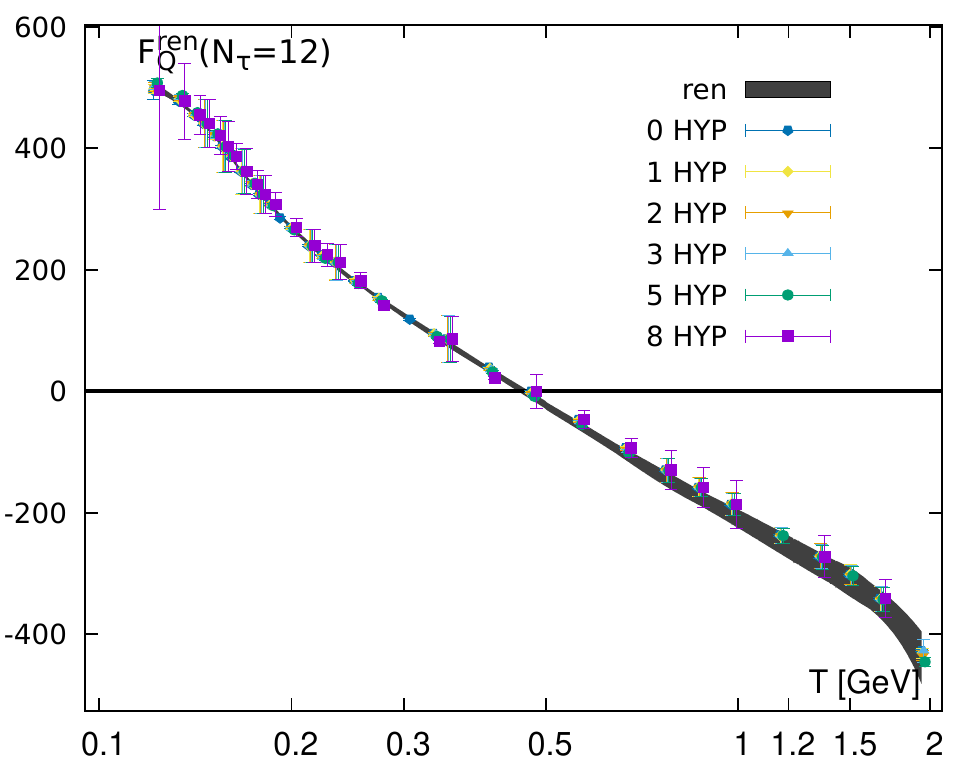}
\caption{The $N_{\tau}$ dependence of $\Delta F_Q^{nm}$ for different 
smearing levels $n$ and $m$ (left) and the temperature dependence of $F_Q$
for $N_{\tau}=12$ for different smearing levels (right). The band shows
the continuum extrapolated unsmeared result.}
\label{fig:DeltaFQ}
\end{figure}
HYP smearing distorts the short distance physics. So it is important to understand
the effects of these distortions of $F_Q$ and the Polyakov loop correlators. First
we would like to understand how the HYP smearing affects the temperature dependence
of $F_Q$. The free energy of a static quark calculated with $n$ steps of HYP smearing is
denoted by $F_Q^n$. In Fig. \ref{fig:DeltaFQ} we show the difference $\Delta F_Q^{nm}=F_Q^n-F_Q^m$
for different number of HYP smearing steps as function of $N_{\tau}$. Smearing changes
the coefficient of the $1/a$ divergence in the free energy of a static quark and therefore,
$F_Q^n$ is different for different $n$.
However, unless
excessive amount of smearing is applied the temperature dependence of $F_Q^n$ should not
depend on the number of smearing steps, $n$. Therefore, $\Delta F_Q^{nm}$ should not depend
on $N_{\tau}$ if cutoff effects are negligible (recall $T=1/(a N_{\tau})$).
From Fig. \ref{fig:DeltaFQ} we see that $\Delta F_Q^{nm}$ is independent of $N_{\tau}$ within
errors for $N_{\tau}>4$. In Fig. \ref{fig:DeltaFQ} we also show the temperature dependence of
$F_Q$ for different levels of smearing and $N_{\tau}=12$. We compare our numerical results
with the continuum extrapolated results on $F_Q$ without smearing
in the standard renormalization scheme \cite{Bazavov:2016uvm}.
The results obtained with different number of HYP smearings steps have been shifted by a constant
to match with the unsmeared results on $F_Q$. We clearly see that the temperature dependence is
not affected by HYP smearing. Thus there is no problem with over-smearing in our study, i.e.
the amount of smearing used is not excessive.

\section{Polyakov loop correlators and screening masses}
We study the Polyakov loop correlators 
\begin{equation}
C_{PL}(r,T)=\langle L(r) \cdot L(0) \rangle,
\end{equation}
as well as the correlators of the real and imaginary parts of the Polyakov loop
\begin{eqnarray}
&
C_{PL}^R(r,T)= \langle {\rm Re} L(r) \cdot {\rm Re} L(0) \rangle, \\[2mm]
&
C_{PL}^I(r,T)= \langle {\rm Im} L(r) \cdot {\rm Im} L(0) \rangle.
\end{eqnarray}
We have $C_{PL}(r,T)=C_{PL}^R(r,T)+C_{PL}^I(r,T)$. 
The main reason to study $C_{PL}^R(r,T)$ and $C_{PL}^I(r,T)$ separately is that
these correlators have quite different large distance behaviors governed by different
screening masses, $m_R$ and $m_I$ \cite{Arnold:1995bh}.
This also can be seen in the purely pertubative picture, where the dominant contribution
to $C_{PL}^R$ comes from the two gluon exchange, while the leading contribution
to $C_{PL}^I$ comes from the three gluon exchange. The perturbative picture 
implies a different behavior of $C_{PL}^R$ and $C_{PL}^I$ also at short distances.
In the weak coupling picture $m_I/m_R=3/2$. Furthermore, at leading order (LO)
$m_R=2 m_D$ and $m_I=3 m_D$, with $m_D$ being the LO chromo-electric Debye mass.
The correlator of the real part of the Polyakov loop can couple to magnetic gluons
\cite{Arnold:1995bh,Braaten:1994qx} and therefore, the long distance behavior of
this correlator can be governed by the magnetic screening mass instead of electric
screening mass. However, for the temperature range of interest the contribution
from electric gluon dominates \cite{Hart:1999dj,Hart:2000ha}.

\begin{figure}
\includegraphics[width=7cm]{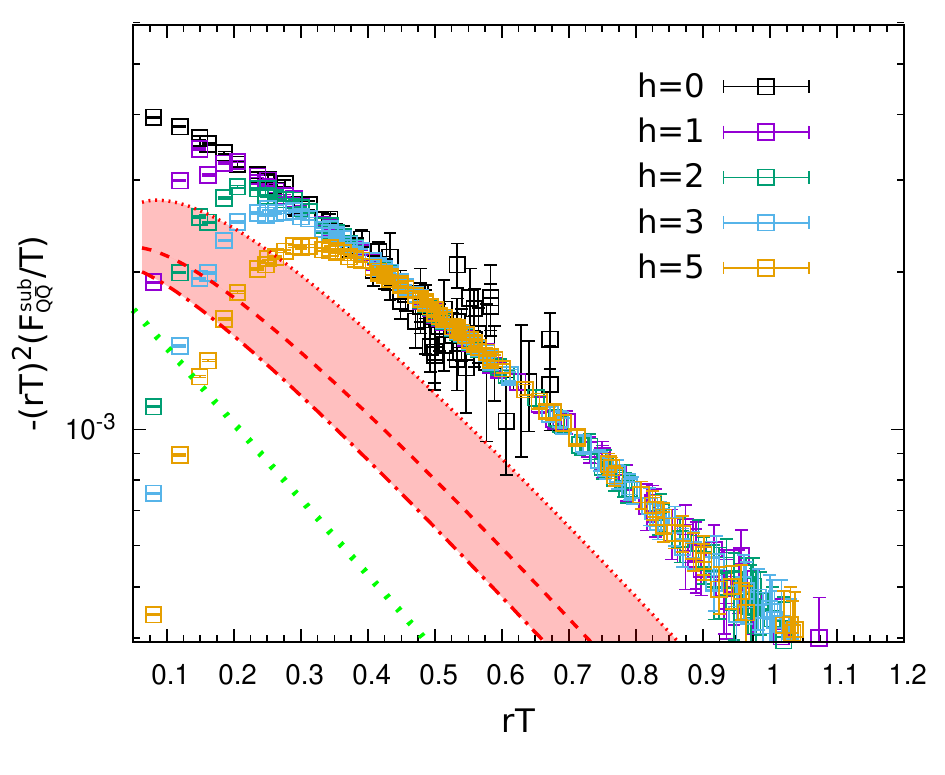}
\includegraphics[width=7cm]{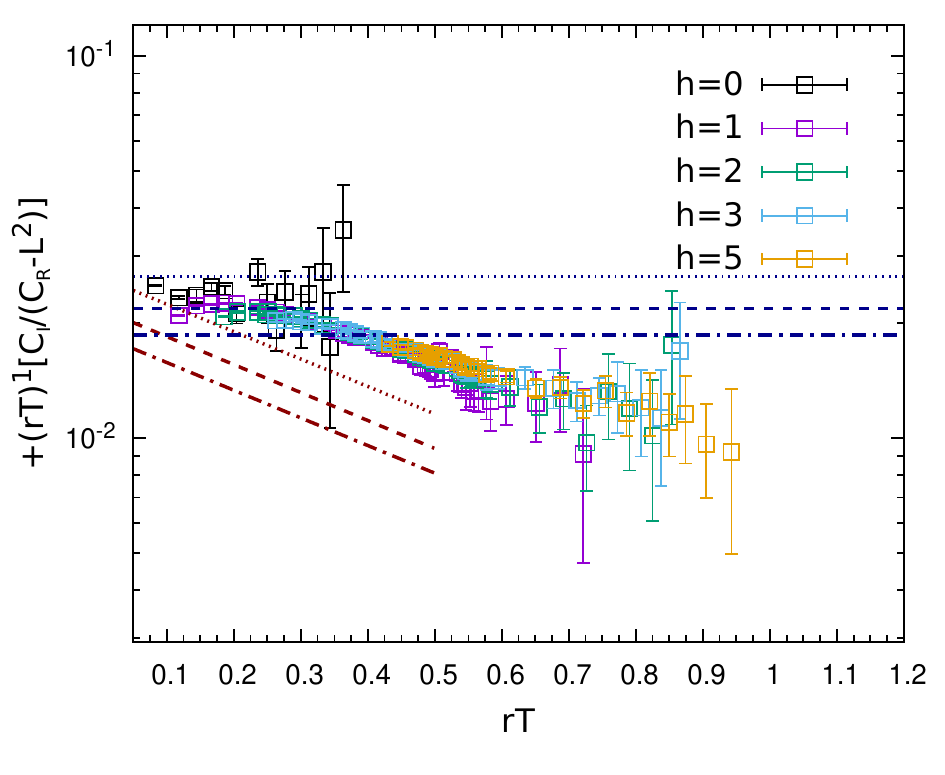}
\caption{The subtracted free energy $F_{Q\bar Q}^{sub}(r,T)$ (left) and 
$R_{PL}^I(r,T)$ (right) as a function of the separation, $r T$ for $N_{\tau}=12$
and $T=1938$ MeV obtained for different number of HYP smearings (h).
In the left panel the LO result (green line) uses renormalization scale $\mu=4\pi T$, 
while for the NLO result (red band) the renormalization scale has 
been varied from $\pi T$ to $4\pi T$.
In the right panel we show the LO result for $R_{PL}^I$ 
for the renormalization scales
$\mu=\pi T,~2 \pi T$ and $4 \pi T$ (from top to bottom). The horizontal lines correspond to
the LO result without screening.}
\label{fig:corr}
\end{figure}

It is convenient to consider the subtracted free energy
\begin{equation}
F_{Q \bar Q}^{sub}(r,T)=-T \ln\frac{C_{PL}(r,T)}{\langle L \rangle^2}=F_{Q \bar Q}(r,T)-2 F_Q(T),
\end{equation}
which is independent of the number of smearing levels used in the calculations except for
small distances. At large distances $F_{Q \bar Q}^{sub}(r,T)$ decays exponentially
with the decay rate given by $m_R$. We also define
the normalized correlators of real and imaginary part of the Polyakov loop 
\begin{eqnarray}
&
\displaystyle
\tilde C_{PL}^R(r,T)= \frac{C_{PL}^R(r,T)-\langle L \rangle^2}{\langle L \rangle^2}, \\[2mm]
&
\displaystyle
\tilde C_{PL}^I(r,T)= \frac{C_{PL}^I(r,T)}{\langle L \rangle^2}.
\end{eqnarray} 
As the subtracted free energy the normalized correlators turn out to be not
very sensitive to the number of smearing steps used in the calculations except
at short distances. The subtracted free energy can be written in terms of
the normalized correlators as
\begin{equation}
F_{Q \bar Q}^{sub}(r,T)/T=\ln(1+\tilde C_{PL}^R(r,T)+\tilde C_{PL}^I(r,T)).
\end{equation}
Since we find that $\tilde C_{PL}^I(r,T) \ll \tilde C_{PL}^R(r,T)$ the subtracted free energy
is dominated by the contribution from $\tilde C_{PL}^R(r,T)$ and can be used as proxy
for $\tilde C_{PL}^R(r,T)$.
Our results for $F_{Q \bar Q}^{sub}(r,T)$ calculated with different numbers of HYP smearings for
$N_{\tau}=12$ lattices and $T=1938$ MeV
are shown in Fig. \ref{fig:corr}. 
The lattice results have been multiplied by $r^2 T$ because of the weak
coupling expectations.
The smeared results are 
compared with the previously calculated unsmeared $F_{Q \bar Q}^{sub}(r,T)$ \cite{Bazavov:2018wmo}.
We see from the figure that even one or two levels of HYP smearing significantly improves the signal compared to
the unsmeared case, however, as we go to larger and larger distances more smearing steps are needed to 
reduce the errors. The smearing clearly alters the behavior of $F_{Q \bar Q}^{sub}(r,T)$  at short distances
but for $rT>0.4$ $F_{Q \bar Q}^{sub}(r,T)$ is independent of smearing level up to three HYP smearing steps.
For five HYP smearing steps the effect of smearing becomes negligible only for $rT >0.6$.
We obtain similar results for other $N_{\tau}$ values and temperatures. The $r/a$ value for which smearing effects 
disappear is independent of $N_\tau$.

We also compare our lattice results with weak coupling expectations. 
In a previous study we showed that the weak
coupling calculations can describe the lattice results of $F_{Q \bar Q}^{sub}(r,T)$ in the short distance region
$rT<0.4$ \cite{Bazavov:2018wmo}. A detailed comparison of the lattice results $F_{Q \bar Q}^{sub}(r,T)$
with the weak coupling calculations at larger distances turned out to be challenging because of the large
statistical errors of the lattice results for $N_{\tau}>4$. 
As discussed in Ref. \cite{Bazavov:2018wmo} at distances $rT>0.4$ 
the appropriate scale hierarchy for the weak
coupling calculation is $r \sim 1/m_D$. The NLO result for this scale hierarchy was obtained in Ref. \cite{Nadkarni:1986cz}
and also confirmed later in Ref. \cite{Berwein:2017thy}.  
In Fig. \ref{fig:corr} we show the comparison of the lattice results with the LO and NLO weak coupling results.
The shape of LO and NLO results is similar to the shape of the lattice results, but
there is no quantitative agreement. While adding the NLO correction moves the weak coupling results
closer to the lattice one, it is still significantly below the lattice result.

In order to study the correlator of the imaginary part of the Polyakov loop we consider the ratio
\begin{equation}
R_{PL}^I=\frac{C_{PL}^I}{C_{PL}^R-\langle L\rangle^2}.
\end{equation}
Again, this ratio is insensitive to smearing. In Fig. \ref{fig:corr} we show it for unsmeared
case as well as for calculations obtained for 1, 2, 3 and 5 steps of HYP smearings for $N_{\tau}=12$
and $T=1938$ MeV. We see that we were able to obtain reasonable results for $R_{PL}^I$ up to distances
$rT \simeq 1$ when 5 steps of HYP smearing are used. Furthermore, the smeared results on $R_{PL}^I$
smoothly connect to the unsmeared results at short distances. In particular, the results with 1 HYP
smearing agree with unsmeared results up to quite small distances. We also compare our lattice
results of $R_{PL}^I$ with the LO results with or without screening for renormalization scales
$\mu=\pi T,~2 \pi T$ and $4 \pi T$, shown as lines. We see that at distances $r T<0.4$ lattice
results agree well with the LO result that does not include screening (horizontal lines). At 
larger distances the lattice data follow the same trend as the LO result with screening but there
are clear quantitative differences. 

\begin{figure}
\includegraphics[width=7cm]{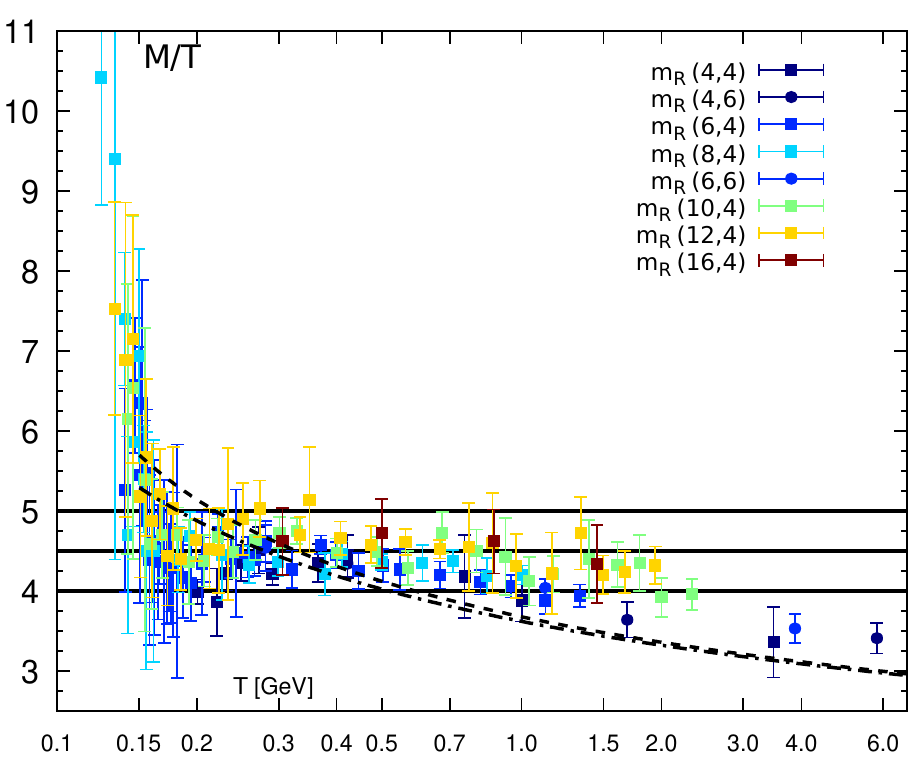}
\includegraphics[width=7cm]{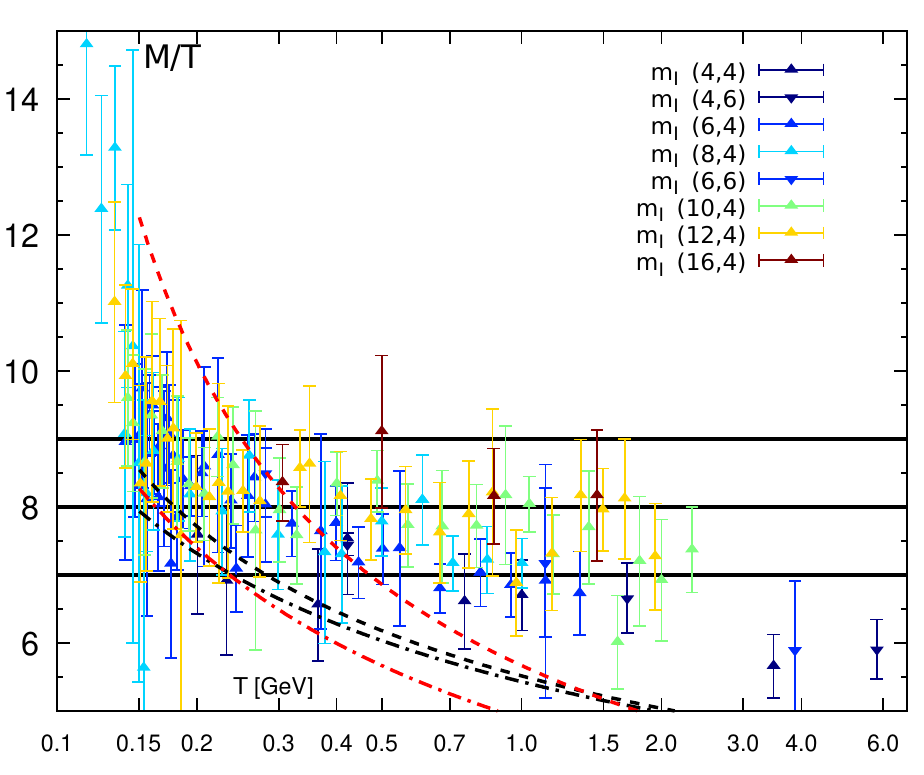}
\caption{The screening masses, $m_R/T$ (left) and $m_I/T$ (right) as a function of the
temperature calculated for different $N_{\tau}$ and $N_{\sigma}/N_{\tau}$ indicated
in the labels by the numbers in the parenthesis.
The solid horizontal lines indicate the approximate value of the screening masses
together with the estimated uncertainty. The dashed lines correspond to the 
perturbative result.}
\label{fig:mscr}
\end{figure}

By fitting the large distance behavior of $\tilde C_{PL}^R(r,T)$ and $\tilde C_{PL}^I(r,T)$
to a $A/r \exp(-m_{R,I} r)$ form, we obtain the corresponding screening masses, $m_R$ and $m_I$.
Our results on the screening masses as functions of the temperature
are shown in Fig. \ref{fig:mscr}. We see a rapid decrease of the screening masses in temperature
units around the crossover temperature, $T_c\simeq 156$ MeV \cite{HotQCD:2018pds},
followed by a roughly constant behavior up to $T \simeq 1.5$ GeV.
For $T<400$ MeV our results for the screening masses are quite similar to the results of 
Ref. \cite{Borsanyi:2015yka}. The decrease of $m_R/T$ and $m_I/T$ predicted by 
the weak coupling calculations is only seen for 
$T> 1.5$ GeV. For the ratio of the screening masses we find $m_I/m_R=1.75$ which is only 17\%
larger than the weak coupling expectation. Therefore, we can define the chromo-electric
screening length as $2/m_R$ or $3/m_I$, which for 170 MeV $<T$ < 1.5 GeV corresponds
to $(0.38-0.44)/T$.

\section{Conclusions}
In this contribution we studied the Polyakov loop and Polyakov loop correlators using HYP
smearing. We found that HYP smearing significantly improves the signal and the distortions
due to smearing can be controlled. The correlator of the imaginary part of the Polyakov loop
is very well described by the leading order perturbative result for $rT <0.4$.
We determined the screening masses of the correlators of the real and imaginary parts of
the Polyakov loop and found $m_R\simeq 4.5 T$ and $m_I \simeq 8 T$, respectively for
 170 MeV $<T$ < 1.5 GeV, implying a chromo-electric screening length of $(0.38-0.44)/T$
in this region.

\acknowledgments
P. Petreczky is supported by the U.S. Department of Energy under Contract No. DE-SC0012704.
S. Steinbei{\ss}er was funded by the Deutsche Forschungsgemeinschaft 
(DFG, German Research Foundation) cluster of excellence 
"ORIGINS" (www.origins-cluster.de) under Germany's Excellence Strategy
EXC-2094-390783311.
J. H. Weber’s research is by the Deutsche Forschungsgemeinschaft (DFG, German Research Foundation) 
Projektnummer 417533893/GRK2575 “Rethinking Quantum Field Theory”. 
The numerical calculations have been performed using the MILC code.
The simulations were carried out on the computing 
facilities of the Computational Center for Particle and Astrophysics (C2PAP) 
in the project ``Calculation of finite T QCD correlators'' (pr83pu).

\bibliographystyle{JHEP}
\bibliography{ref.bib}

\end{document}